\documentclass[runningheads]{llncs}
\usepackage{graphicx}
\usepackage[utf8]{inputenc}

\usepackage{color}
\usepackage{amsmath}
\usepackage{hyperref}

\DeclareUnicodeCharacter{2212}{-}

\begin{document}

\title{Improved inter-scanner MS lesion segmentation by adversarial training on longitudinal data 
       \thanks{This project received funding from the European Union's Horizon 2020 research and innovation program under the Marie Sklodowska-Curie grant agreement No 765148. The computational resources and services used were provided by the VSC (Flemish Supercomputer Center), funded by the Research Foundation - Flanders (FWO) and the Flemish Government - department EWI. The final authenticated publication is available online at \url{https://doi.org/10.1007/978-3-030-46640-4_10}.}
        }

\titlerunning{Improved inter-scanner MS lesion segmentation}

\author{Mattias Billast\inst{1} \and 
        Maria Ines Meyer \inst{2,3} \and 
        Diana M. Sima \inst{2} \and 
        David Robben \inst{2,4} }

\authorrunning{M. Billast, M. I. Meyer, D. Sima, D. Robben }
%

\institute{
          Dept. of Mathematical Engineering, KU Leuven, Belgium \\ \email{mattias.billast@gmail.com} 
          \and icometrix, Leuven, Belgium  \\
          \email{ \{ ines.meyer, diana.sima,david.robben \} @icometrix.com}
          \and Dept. of Health Technology, Technical University of Denmark, Lyngby, Denmark \\ 
          \and Medical Image Computing (ESAT/PSI), KU Leuven, Belgium
}


\maketitle              

\begin{abstract} 
The evaluation of white matter lesion progression is an important biomarker in the follow-up of MS patients and plays a crucial role when deciding the course of treatment. Current automated lesion segmentation algorithms are susceptible to variability in image characteristics related to MRI scanner or protocol differences.  We propose a model that improves the consistency of MS lesion segmentations in inter-scanner studies. First, we train a CNN base model to approximate the performance of ico\textbf{brain}, an FDA-approved clinically available lesion segmentation software. A discriminator model is then trained to predict if two lesion segmentations are based on scans acquired using the same scanner type or not, achieving a $78\%$ accuracy in this task. Finally, the base model and the discriminator are trained adversarially on multi-scanner longitudinal data to improve the inter-scanner consistency of the base model. The performance of the models is evaluated on an unseen dataset containing manual delineations. The inter-scanner variability is evaluated on test-retest data, where the adversarial network produces improved results over the base model and the FDA-approved solution.

\end{abstract}

\section{Introduction}

Multiple sclerosis (MS) is an autoimmune disorder characterized by a  demyelination process which results in neuroaxonal degeneration and the appearance of lesions in the brain. The most prevalent type of lesions appear hyperintense on T2-weighted (T2w) magnetic resonance (MR) images and their quantification is an important biomarker for the diagnosis and follow-up of the disease \cite{RN3}.

Over the  years methods for automated lesion segmentation have been developed. Several approaches model the distribution of intensities of healthy brain tissue and define outliers to these distributions as lesions \cite{VanLeemput2001,Jain2015}. Others are either atlas-based \cite{Shiee2010} or data-driven (supervised) \cite{Brosch2016,Valverde2017} classifiers. For a detailed overview of recent methods refer to \cite{RN3}. 

\begin{figure}[t] 
    \setlength{\belowcaptionskip}{-12pt}
    \centering
    \includegraphics[width=0.7\textwidth]{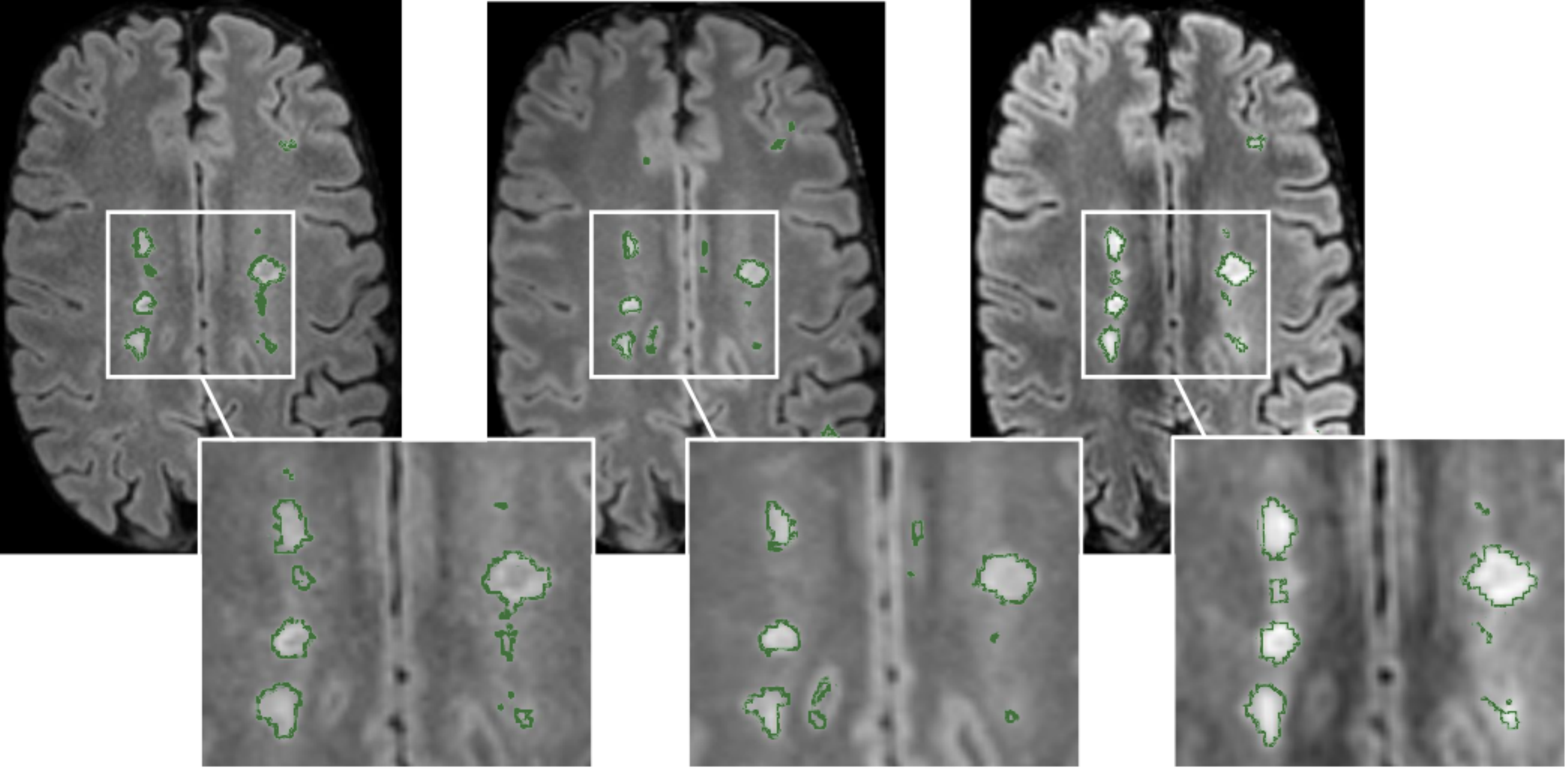}
    \caption{MRI scans from one patient in three 3T scanners (left to right: Philips Achieva, Siemens Skyra and GE Discovery MR750w). Automated lesion segmentations in green.}
    \label{fig:scan_comparison}
\end{figure}

Lesion segmentation is particularly interesting for patient follow-up, where data from two or more time-points is available for one patient. Some approaches try to improve segmentation consistency by analysing intensity differences over time \cite{JainS.2017Uffc}.
Although these methods achieve good performance in controlled settings, they remain sensitive to changes in image characteristics related to scanner type and protocol. In a test-retest multi-scanner study \cite{Biberacher2016}, scanner type was observed to have an effect on MS lesion volume. These findings are supported by \cite{Shinohara2017}, where scanner-related biases were found even when using a harmonized protocol across scanners from the same vendor. Fig. \ref{fig:scan_comparison} illustrates such an effect.

Few works have addressed the inter-scanner variability issue in the context of lesion segmentation. 
Recent approaches attempt to increase the generalization of CNN-based methods to unseen MR scanner types through domain adaptation \cite{Kamnitsas2016} or transfer learning \cite{Ghafoorian2017,Valverde2019} techniques.
Nevertheless, these methods share the common downside that they require a training step to adapt to new unseen domains (scanners types and protocols). The consistency of the delineations in longitudinal settings is also not considered. A solution to incorporate consistency information into this type of data-driven solutions  would be to train them on a dataset containing intra- and inter-scanner repetitions for the same patient, acquired within a short periods of time. However, in practice this type of \emph{test-retest} dataset is almost impossible to acquire at a large scale, due to time and cost considerations.

In the present work we present a novel approach to improve the consistency of lesion segmentation in the case of multi-scanner studies, by capturing inter-scanner differences from lesion delineations. Given the shortage of test-retest data we propose instead to use longitudinal inter-scanner data to train a cross-sectional method. We start by training a base model on a multi-scanner dataset to achieve performance comparable to an existing lesion segmentation software \cite{Jain2015}. We then design a discriminator to identify if two segmentations were generated from images that originate from the same scanner or not.  The assumption is that the natural temporal variation in lesion shape can be distinguished from the variation caused by the different scanners. These networks are then combined and trained until the base model produces segmentations that are similar enough to fool the discriminator. We hypothesize that through this training scheme the model will  become invariant to scanner differences, thus imposing consistency on the baseline CNN.
Finally we evaluate the accuracy on a dataset with manual lesion segmentations and the reproducibility on a multi-scanner test-retest dataset.

\section{Methods}

We start by building a lesion segmentation \emph{base model} based on a deep convolutional neural network (CNN) architecture \cite{Kamnitsas2017} that approximates the performance of ico\textbf{brain}, an FDA-approved segmentation software. 
This method is an Expectation-Maximization (EM) model that uses the distribution of healthy brain tissue to detect lesions as outliers while also using prior knowledge of the location and appearance of lesions \cite{Jain2015}. We refer to it as \emph{EM-model}.

\subsubsection{Base Model}
\begin{figure}[t]
\setlength{\belowcaptionskip}{-10pt}
    \centering
    \includegraphics[clip, trim=0.5cm 3cm 0.5cm 3cm, width=1.0\textwidth]{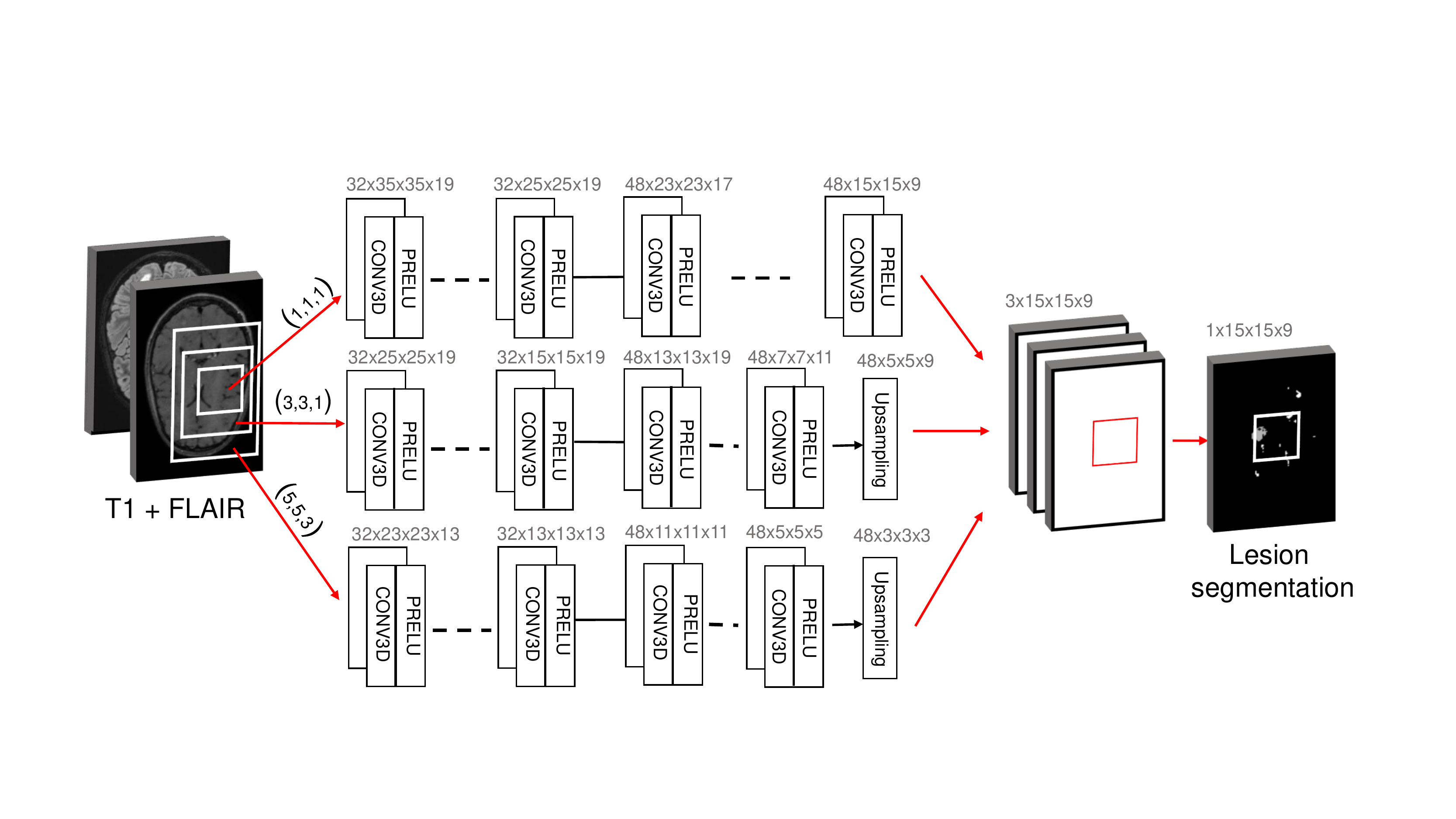}
    \caption{Architecture of the base model that describes the patch sizes of the different pathways and the overall structure.}
    \label{fig:base_model}
\end{figure}

The base model is based on the DeepMedic architecture \cite{Kamnitsas2017}. Generally, it is composed of multiple pathways which process different scales of the original image simultaneously. This is achieved by downsampling the original image at different rates before dividing it into input patches, which allows the model to combine the high resolution of the original image and the broader context of a downsampled image to make a more accurate prediction. In our implementation we used three pathways, for which the input volumes were downsampled with factors $(1,1, 1)$, $(3, 3, 1)$ and $(5, 5, 3)$ and divided into patches of size $(35,35,19)$, $(25, 25, 19)$ and $(23, 23, 13)$, respectively. 
Each pathway is comprised of ten convolutional layers, each followed by a PReLu activation, after which the feature maps from the second and third pathways are upsampled to the same dimensions as the first pathway and concatenated. This is followed by dropout, two fully connected layers and a sigmoid function, returning a $(15,15,9)$ probability map. The first five layers have 32 filters and kernel size (3,3,1) and the last five layers 48 filters with kernel size (3,3,3). The values of the output probability map that are above a certain threshold are classified as lesions. The threshold used throughout this article is 0.4. The architecture is represented in Fig. \ref{fig:base_model}.
The loss function of the base model is given by
\begin{equation}
     L_B =  Ylog(B(X)) + (1-Y)log(1-B(X)),
     \label{eq:kost}
\end{equation}
where X is the concatenation of the T1- and FLAIR MR images, Y is the corresponding lesion segmentation label and B() the output of the base model.

\subsubsection{Discriminator}
\begin{figure}[t]
\setlength{\belowcaptionskip}{-10pt}
    \centering
    \includegraphics[clip, trim=1.5cm 5.5cm 1.5cm 4cm, width=1.0\textwidth]{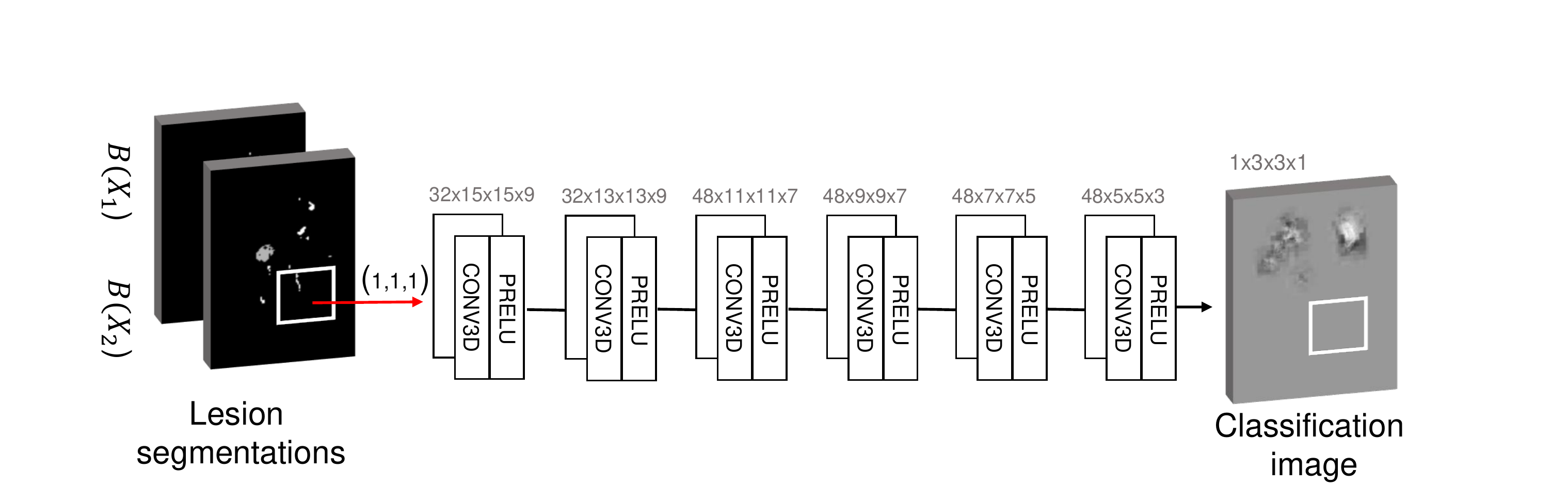}
    \caption{Architecture of the discriminator that describes the in- and output sizes of the patches and the overall structure.}
    \label{fig:discriminator}
\end{figure}

The discriminator is reduced to one pathway with six convolutional layers, since additional pathways with subsampling resulted in a marginal increase in performance. The two first layers have 32 filters of kernel size (3,3,1) and the following layers 48 filters with kernel size (3,3,3). As input it takes two label patches of size $(15,15,9)$ and generates a voxel-wise prediction that the two labels are derived from images acquired using the same scanner. The architecture is represented in Fig. \ref{fig:discriminator}.

The loss function of the discriminator is given by :
\begin{equation}
     L_D =  Ylog(D(B(X_1),B(X_2))) + (1-Y) log(1-D(B(X_1),B(X_2))),
     \label{eq:kost}
\end{equation}
where $Y$ is the ground truth indicator variable (0 or 1) indicating whether two time points were acquired on the same scanner or not, $X_1$ and $X_2$ are images at different time points and B() and D() are respectively the output of the base model and the discriminator.

\subsubsection{Adversarial Model}
\begin{figure}[t]
    \centering
    \includegraphics[width=0.7\textwidth]{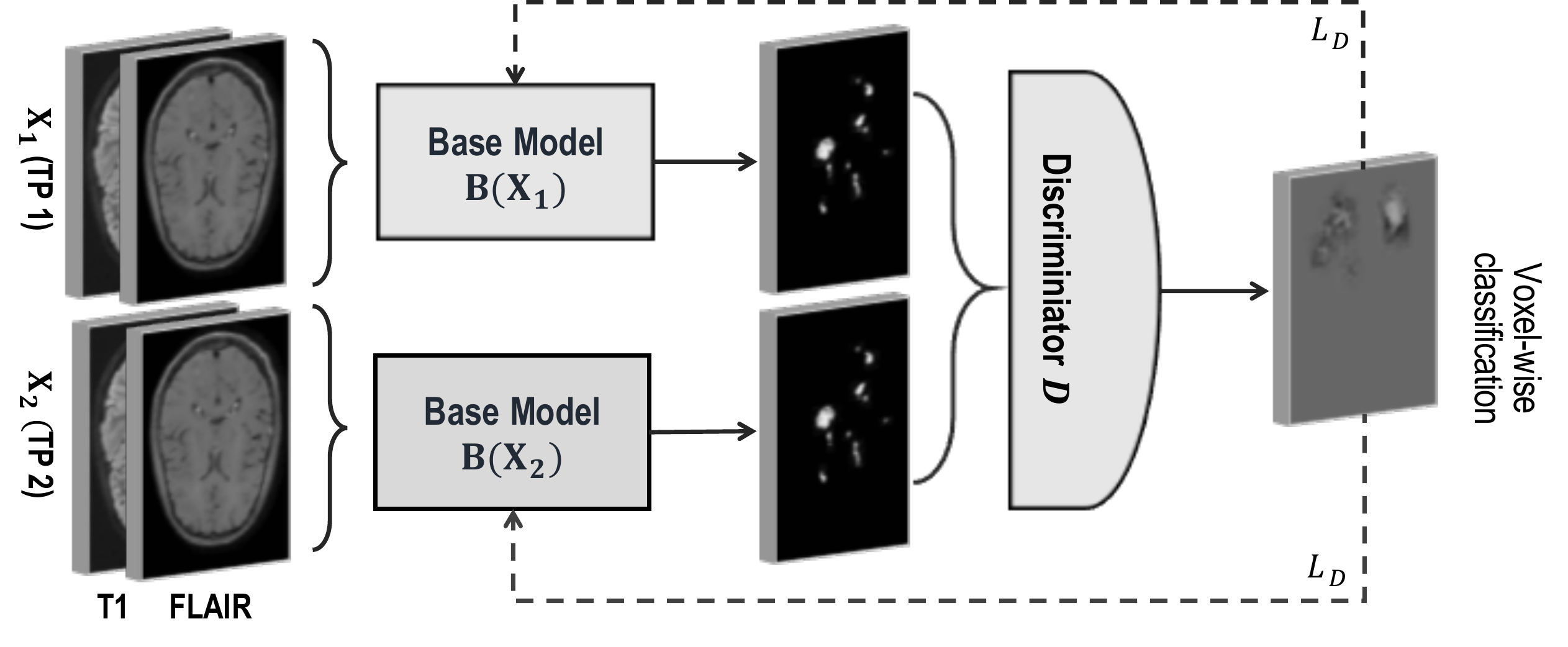}
    \caption{Adversarial network that combines the base model and the discriminator to reduce the inter-scanner variability.}
    \label{fig:adv_net}
\end{figure}

After training, the discriminator was combined adversarially with the base model, as introduced in \cite{GoodfellowIanJ.2014GAN}. The adversarial model consists of two base model blocks ($B$) and one discriminator ($D$) (Fig. \ref{fig:adv_net}). 
In our particular case the pre-trained weights of the discriminator are frozen and only the weights of the base model are fine-tuned. The concept of adversarial training uses the pre-trained weights of the discriminator to reduce the inter-scanner variability of the base model by maximizing the loss function of the discriminator. This is equivalent to minimizing the following loss function : 
\begin{equation}
     L_{Adv} =  (1-Y)log(D(B(X_1),B(X_2))) + Ylog(1-D(B(X_1),B(X_2))),
     \label{eq:kost1}
\end{equation}
The loss function of the adversarial network then consists of two terms: one associated with the lesion segmentation labels, and one related to the output image of the discriminator:
\begin{equation}
     L =  2*L_B + L_{Adv}
     \label{eq:kost2}
\end{equation}

The purpose of $L_{Adv}$ is to ensure that the base model is updated such that the discriminator can no longer distinguish between segmentations that are based on same- or different-scanner studies. We hypothesize that the base model learns to map scans from different scanners to a consistent lesion segmentation.

\subsubsection{Model training}
Both the base model and the discriminator were trained using the binary cross entropy objective function and optimized using mini-batch gradient descent with Nesterov momentum $\beta=0.9$. Initial learning rates were $\alpha=0.016$ for the base model and  $\alpha=4\mathrm{e}{-3}$ for the discriminator, and were decreased at regular intervals until convergence. For the adversarial network initial learning rate was $\alpha=2\mathrm{e}{-3}$. 
All models were trained using an NVIDIA P100. The networks are implemented using the Keras and DeepVoxNet \cite{RobbenDavid2018Dvpf} frameworks.

\section{Data and preprocessing} \label{sec:data}
Four different datasets were available: two for training and two for testing the performance of the models. Since for three of the datasets manual delineations were not available, automated segmentations were acquired using the EM-method described in the previous section. All automated delineations were validated by a human expert. Each study in the datasets contains T1w and FLAIR MR images from MS patients.\\
\emph{\textbf{Cross-sectional dataset}} 208 independent studies from several centers. The base model is trained on this dataset.\\
\emph{\textbf{Longitudinal dataset}} 576 multi-center, multi-scanner studies with approved quality MR scans, containing multiple studies from 215 unique patients at different timepoints. For training the adversarial model and the discriminator only studies with less than 2 years interval were used to minimize the effect of the natural evolution of lesions over time and capture the differences between scanners.
This resulted in approximately $80\%$ being used since most patients have a follow-up scan every 6 months to one year. The discriminator and adversarial model are trained on this dataset. \\
\emph{\textbf{Manual segmentations}} 20 studies with manual lesion delineations by experts. \\
\emph{\textbf{Test-retest dataset }} 10 MS patients. Each patient was scanned twice in three 3T scanners: \emph{Philips Achieva}, \emph{Siemens Skyra} and \emph{GE Discovery MR450w} \cite{Jain2015}. \\

All the data was registered to Montreal Neurological Institute (MNI) space and intensities were normalized to zero mean and unit standard deviation. Ten studies from each training dataset were randomly selected to use as validation during the training process. The data was additionally augmented by randomly flipping individual samples around the x-axis.

\section{Results}

\begin{figure}[t]
    \centering
    \includegraphics[width=0.9\textwidth]{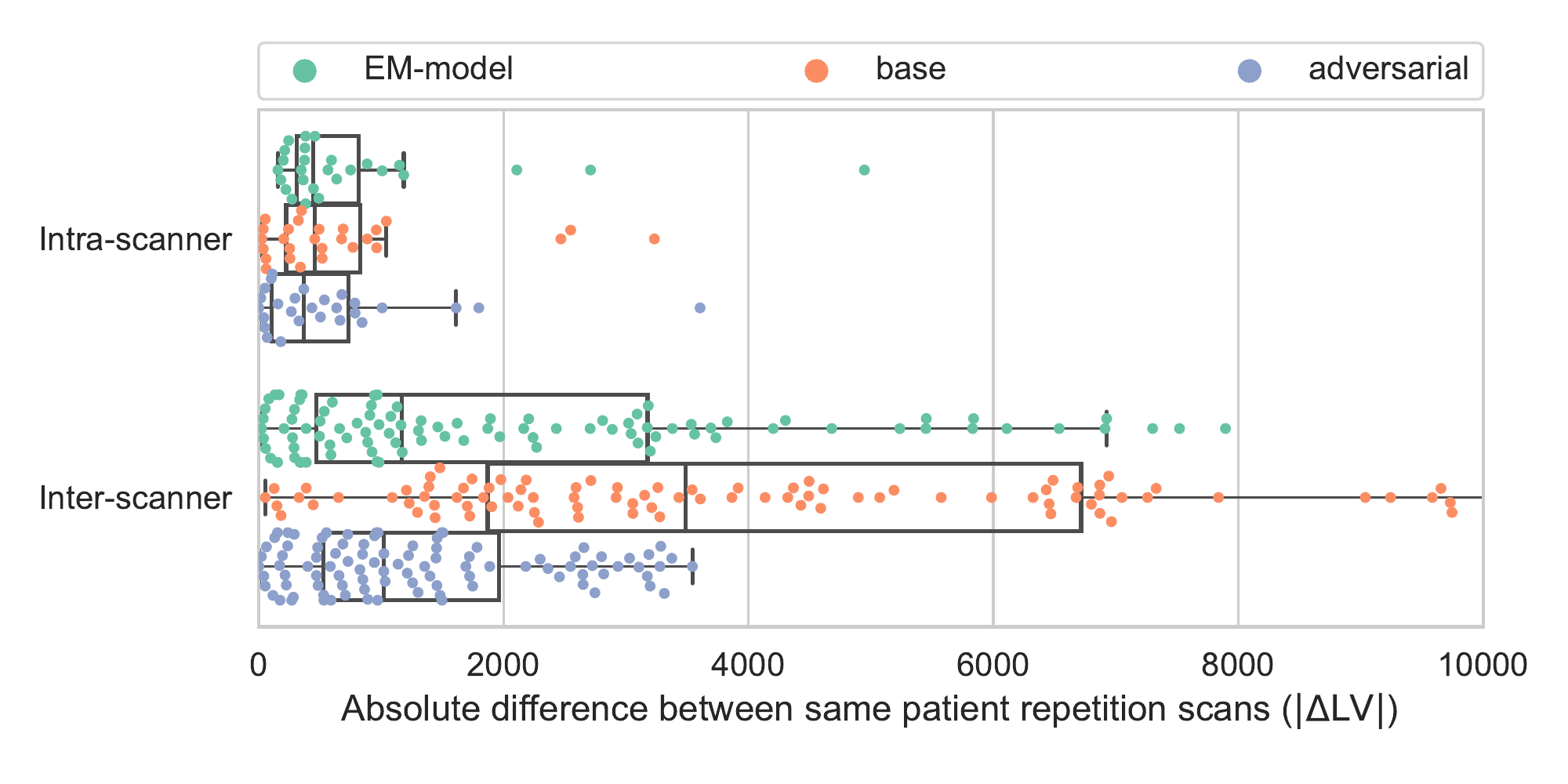}
    \caption{Absolute intra- and inter-scanner difference in lesion volume, calculated on the test-retest dataset with three different models.}
    \label{fig:model_comparison}
\end{figure}

The models were evaluated on the manual segmentations and the test-retest datasets described in Section \ref{sec:data} and compared to the EM-model.
The main results are summarized in Table \ref{tab:results}. For the manual segmentations dataset results are described in terms of Dice score, Precision and Recall.
For the test-retest dataset we are mainly interested in evaluating the reproducibility in the inter-scanner cases. Since there is no ground truth, we report the metrics between different time points for the same patient. Aside from the total lesion volume (LV) in $mm^3$ we additionally quantify the absolute differences in lesion volume ($|\Delta \mathrm{LV}|$) in $mm^3$. 
The results in this table  were calculated with a lesion threshold value of $0.4$. 
Fig. \ref{fig:model_comparison} depicts the distribution of ($|\Delta \mathrm{LV}|$) for both inter-scanner and intra-scanner cases of the test-retest dataset. 

\paragraph{\textbf{Base model}}
For the manual segmentation dataset, results are comparable to the EM-model. 
In the test-retest validation, the inter-scanner $|\Delta \mathrm{LV}|$ is larger for the base model, which indicates that the model is sensitive to inter-scanner variability. 

\paragraph{\textbf{Discriminator}} The discriminator is validated on a balanced sample of the test-retest dataset, so that there is the same number of inter- and intra-scanner examples. It achieves an accuracy of $78\%$ by looking at the average probability value on the lesion voxels only.

\paragraph{\textbf{Adversarial model}}
On the manual segmentations dataset, again referring to Table \ref{tab:results}, the adversarial model achieves a slightly lower but still competitive performance when compared to the EM-model.

Regarding the test-retest dataset, the adversarial model produces lower inter-scanner $|\Delta \mathrm{LV}|$ when compared to the base model (Wilcoxon Signed-Rank Test, $\mathrm{p} = 3.26e-15)$ and to the EM-model, (Wilcoxon Signed-Rank Test, $\mathrm{p} = 0.02$).
This indicates that the adversarial model produces segmentations that are less sensitive to inter-scanner variation than both the base model and the EM-model.

The mean $|\Delta \mathrm{LV}|$ values and standard deviation for the EM-model are almost twice as large as the adversarial model. Taking into account the boxplots in Fig. \ref{fig:model_comparison}, this is partly explained by the fact that the distribution has a positive skew and additionally by three significant outliers, which artificially increase the mean values.  

\begin{figure}[t]
    \centering
    \includegraphics[width=0.9\textwidth]{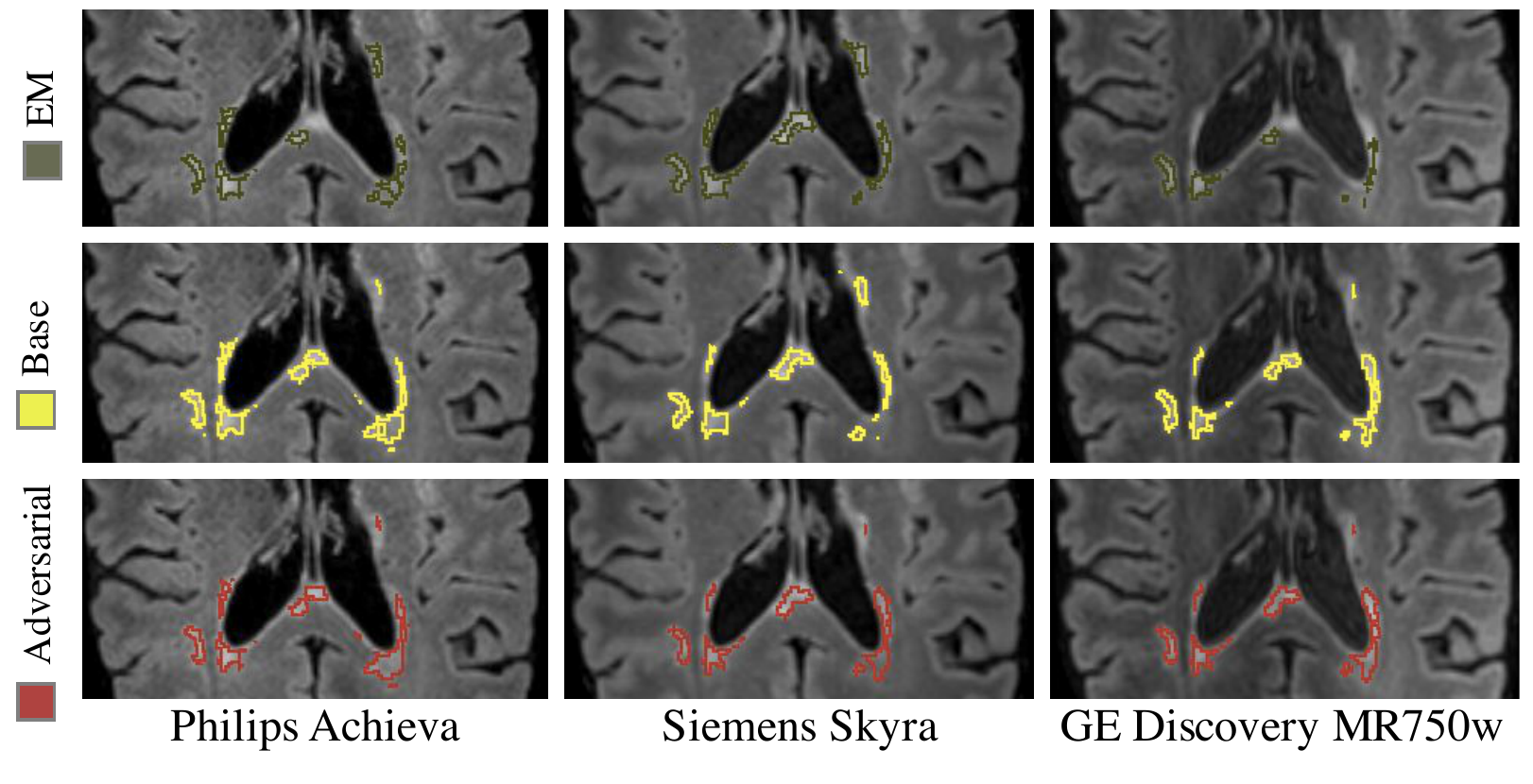}
    \caption{Lesion segmentation results for one patient in three 3T scanners. \textbf{Top:} EM-model; \textbf{Middle:} base model; \textbf{Bottom:} adversarial model. Adversarial model results appear more consistent, while maintaining physiological meaning.}
    \label{fig:segmentations_comparison_models}
\end{figure}

This is evidence that the EM-model has larger variability and lower reproducibility than the adversarial model, while the average predicted lesion volume is similar for the EM- and adversarial models. Fig. \ref{fig:segmentations_comparison_models} shows an example of the different lesion segmentations on the different scanners with the three models.

\setlength{\tabcolsep}{5pt}
\begin{table}[t]
    \caption{Mean performance metrics for the different models on two test sets: manual segmentations and test-retest. For the latter only inter-scanner studies are considered. $|\mathrm{LV}|$ represents absolute differences between individual lesion volumes and is given in $mm^3$.}
    \centering
    \begin{tabular}{c|ccc||cc}
         & \multicolumn{3}{c||}{Manual} & \multicolumn{2}{c}{Test/Retest} \\ 
         Model      & Dice              &  Precision   & Recall     & $|\Delta \mathrm{LV}|$  & LV   \\ \hline
         EM         & $0.71 \pm 0.07$   & $0.85$    & $0.61$        & $2077 \pm 2054$  &  $8307$\\
         Base       & $0.72 \pm 0.10$   & $0.80$    & $0.65$        & $4557 \pm 3530$  &  $9894$\\
         Adversarial& $0.68 \pm 0.11$   & $0.83$    & $0.59$        & $1331 \pm 1020$  &  $8584$\\
    \end{tabular}
    \label{tab:results}
\end{table}

\section{Discussion and future work}

We presented a novel approach to improve the consistency of inter-scanner MS lesion segmentations by using adversarial training on a longitudinal dataset. The proposed solution shows improvements in terms of reproducibility when compared to a base CNN model and to an FDA-approved segmentation method based on an EM approach. The key ingredient in the model is the discriminator, which predicts with $78\%$ accuracy on unseen data whether two lesion segmentations are based on MRI scans acquired using the same scanner. This is a very promising result, since this is not a standard problem.

When evaluated on an unseen dataset of cross-sectional data, the model's performance approximates the EM-model, but decreases slightly after the adversarial training. This indicates a trade-off between performance and reproducibility. One concern was that this would be connected to an under-segmentation due to the consistency constraint learned during the adversarial training. However, evaluating the average predicted lesion volume on a separate test-retest dataset shows no indication of under-segmentation when compared to the EM-model.

Both the adversarial network and the discriminator were trained on longitudinal inter-scanner data. This is not ideal, since MS can have an unpredictable evolution over time, and as such it becomes difficult to distinguish between differences caused by hardware and the natural progression of the disease. We attempt to mitigate this effect by selecting studies within no more than two years interval, but better and more reliable performance could be achieved if the model would be trained on a large dataset  with the same characteristics as the test-retest dataset described in section \ref{sec:data}. However, large datasets of that type do not exist and would require a very big effort to collect, both from the point of view of patients and logistics. As such, using longitudinal inter-scanner data is a compromise that is cost-efficient and shows interesting results.

Another point that could improve the performance would be to use higher quality images and unbiased segmentations at training time. This would allow for a stronger comparison to other methods in literature and manual delineations. At this moment it is expectable that our model achieves results comparable to those of the method used to obtain the segmentations it was trained on.

Aside from these compromises, some improvements can still be made in future work. 
Namely, during the training and testing stages of the adversarial network images can be affinely registered to each other instead of using one common atlas space. We would expect this to increase the overlap metrics.
On the other hand it was observed that the overlap metrics slightly decrease for the adversarial network with longer training, and as such the weight of the term in the loss function associated with the discriminator can be optimized/lowered to achieve more efficient training and better overlap of the images.

Finally, instead of only freezing the weights of the discriminator to improve the base model, the weights of the base model can also be frozen in a next step to improve the discriminator, so that the base model and discriminator are trained in an iterative process until there are no more performance gains.

Apart from the various optimizations to the model, it would be interesting to apply the same adversarial training to other lesion types, such as the ones resulting from vascular dementia or traumatic brain injuries.

%
%


%
%
%
\bibliographystyle{splncs04}
\bibliography{references}

\end{document}